\documentclass[showpacs,twocolumn,aps,prb]{revtex4-1}
\usepackage{amssymb}
\usepackage{amsmath}
\usepackage{graphicx}
\usepackage{subfigure}
\usepackage{color}
\usepackage{ulem}
\usepackage{bm}
\usepackage{makecell}
\usepackage{makecell}
\usepackage{multirow}
\usepackage{diagbox}
\usepackage{booktabs}

\begin{document}
\title{The unidirectional Seebeck detection of the N\'{e}el vector in the two-dimensional tetragonal $\mathcal{PT}$-symmetric antiferromagnetic materials}

\author{Ya-Ting Xiao$^{1}$}
\altaffiliation{These authors contributed equally to this work.}
\author{Ying-Li Wu$^{1}$}
\altaffiliation{These authors contributed equally to this work.}
\author{Jia-Liang Wan$^{1}$}

\author{Xiao-Qin Yu$^{1}$}
\email{yuxiaoqin@hnu.edu.cn}

\affiliation{$^{1}$ School of Physics and Electronics, Hunan University, Changsha 410082, China.}

\begin{abstract}
The efficient detection of the reversal (180$^{\circ}$ rotation) of the  N\'{e}el vector is one of the crucial tasks in antiferromagnetic spintronics. Here, we propose a thermal approach to detect the reversal of the N\'{e}el vector in the tetragonal $\mathcal{PT}$ antiferromagnetic materials through the unidirectional Seebeck effect (USE). Being different from the previous works in which USE stems from the global Rashba spin-orbit coupling (SOC) or asymmetric magnon scattering, we find that the USE originates from the coupling of the hidden Rashba SOC and the N\'{e}el vector in the tetragonal $\mathcal{PT}$ antiferromagnetic materials in the absence of the global Rashba SOC. Using a generic minimal model, we analyse the behaviors of the USE for the two-dimensional tetragonal lattice $\mathcal{PT}$ antiferromagnet.
Importantly, It's found that when the N\'{e}el vector is reversed, the sign of the USE changes, which can be utilized to detect the reversal of the N\'{e}el vector.

\end{abstract}

\maketitle
\section{Introduction}
\label{Introduction}
Antiferromagnets (AFMs),
owing to their insensitivity to magnetic perturbations, zero stray field, and ultra-fast dynamics,
display a potential to complement or replace ferromagnets as the active spin-dependent element of spintronics devices and lead to the emergence of a tremendously active subfield of spintronics: antiferromagnetic spintronics\cite{Baltz,Jungwirth,Han,Jungwirth2}. Using AFMs as active
data storage media, one of the crucial tasks that antiferromagnetic spintronics face is the efficient detection of the reversal ($180^{\circ}$ rotation) of the N\'{e}el vector, namely
the antiferromagnetic order parameter\cite{Baltz,Jungwirth}.
Because of the lack of a net magnetization in AFMs, the conventional magnetic techniques are useless to detect the reversal of the N\'{e}el vector, although the anisotropic magnetoresistance effect (a widely employed conventional magnetic technique) has been  utilized to detect the reorientation of the N\'{e}el vector but only limits to the 90$^{\circ}$ rotation of the N\'{e}el vector\cite{Zelezny}. 

Recently, several electrical approaches have been proposed to detect the reversal of N\'{e}el vector, including intrinsic nonlinear Hall effect in three-dimensional $\mathcal{PT}$-symmetric compensated AFMs\cite{Huiying,C.Wang} and two-dimensional (2D) AFMs\cite{J.Wang}, the in-plane anomalous Hall effect in $\mathcal{PT}$-symmetric antiferromagnetic Materials\cite{J.Cao}, and the layer Hall effect in centrosymmetric magnetoelectric AFMs \cite{L.L.Tao}. In these approaches, the relied-upon electrically driven unconventional transverse effects (i.e., Hall-like effects) all originate from the geometric properties of energy bands (such as quantum metric, Berry curvature, and hidden Berry curvature).

\begin{figure*}
\centering
\begin{minipage}[b]{0.66\linewidth}
 \includegraphics[width=1.0\textwidth]{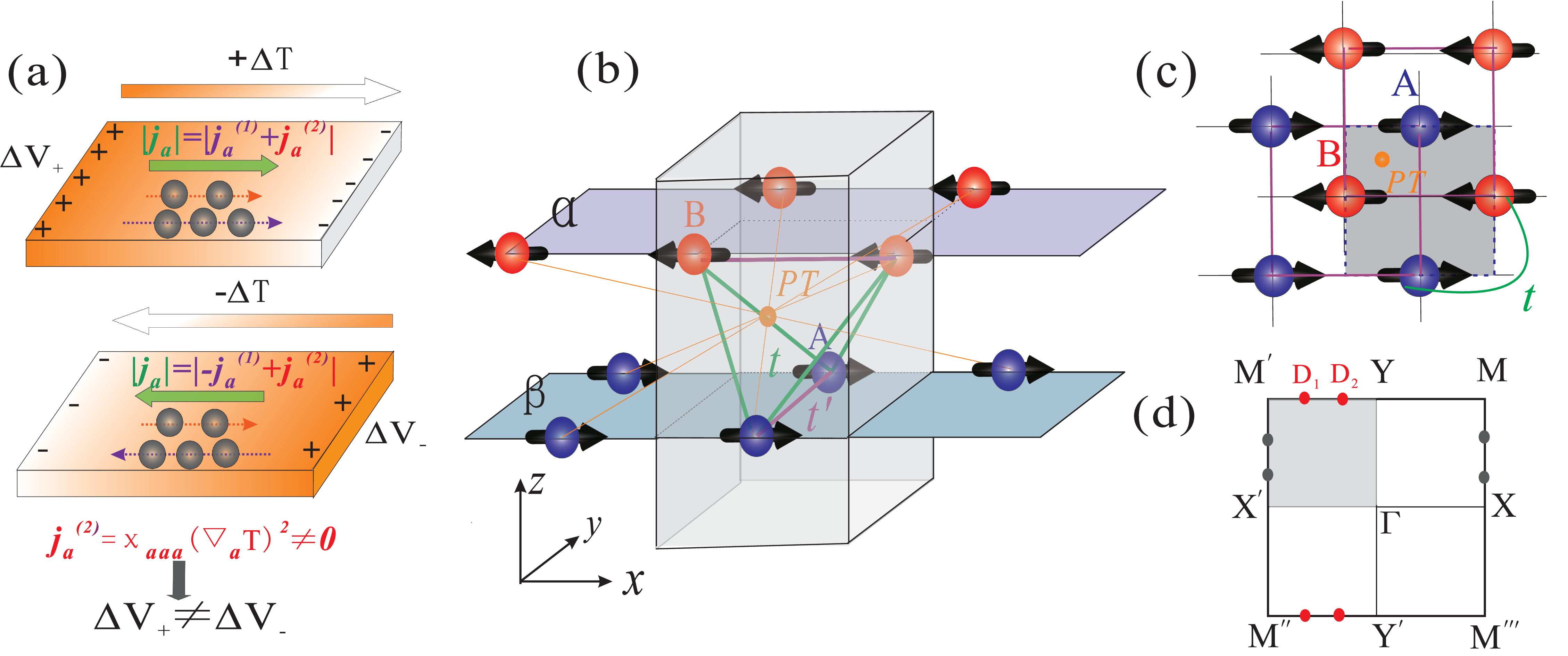}
\end{minipage}
\begin{minipage}[b]{0.315\linewidth}
 \includegraphics[width=1.0\textwidth]{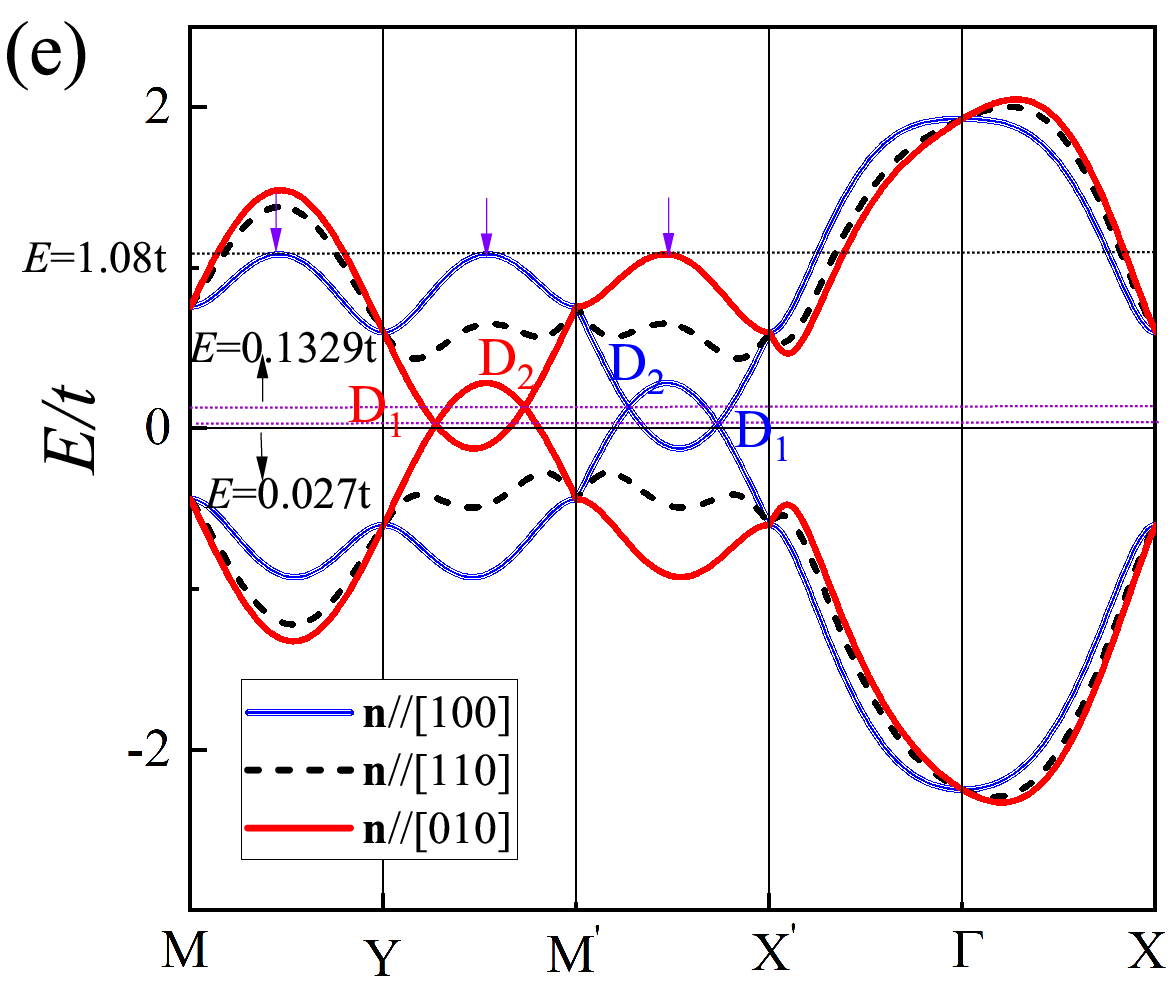}
\end{minipage}
\caption{(a) Illustration of the concept of USE under $+\Delta T$ (top) and $-\Delta T$ (bottom) temperature difference for a positive $\chi_{aaa}$ case. $j_{a}^{(1)}\propto (\nabla_{x}T)$ ($j_{a}^{(2)}=\chi_{aaa}(\nabla_{x}T)^{2}$) are currents as a  first- (second-)order response to temperature gradient. For the negative $\chi_{aaa}$ case, it is similar except the direction of $j_{a}^{(2)}$ is reversed. (b) The crystal structure of the tetragonal lattice $\mathcal{PT}$ antiferromagnet is composed of two sublattices A and B (or say two sectors, $\alpha$ and $\beta$) connected by the $\mathcal{PT}$-symmetric center marked by the original ball. The gray box shows the unit cell. (c) Top view of the quasi-2D-antiferromagnet model. The green (purple) lines denote the first (second) nearest-neighbor hopping $t$ ($t^{\prime}$). (d) The 2D Brillouin zone (BZ) projection with the Dirac point (DP) positions. The black (red) points indicate the DPs for $n\parallel [100]$ ($n\parallel [010]$). (e) The energy band dispersion of the considered generic minimal model along $M-Y-M^{\prime}-X^{\prime}-\Gamma-X$ for three different orientations of in-plane N\'{e}el vector. The parameters are taken as follows: $t^{\prime}=0.08t$, $\lambda=0.8t$, and $j_{n}=0.6t$.}
\label{figure1}
\end{figure*}

In addition to the unconventional nonlinear transverse effects, the recently observed unidirectional magnetoresistance\cite{Avci-2015,Yasuda,Lv,He,He2,Ideue}, a nonlinear longitudinal effect, has attracted broad interest to the study of nonlinear longitudinal effects independent of the band geometry, such as second-order magnetoresistance\cite{Godinho-2018}, bilinear magnetoresistance\cite{Dyrdal,Wang-2022}, and unidirectional Seebeck effect (USE)\cite{Yu-2019,Zhu-2023}. It has been shown that the second-order magnetoresistance can also be applied to electrically detect the reversal of the N\'{e}el vector \cite{Godinho-2018}. Being different to the other three nonlinear longitudinal effects driven by the electric field, USE refers to a thermally driven nonlinear magnetothermal phenomenon, in which the magnitude of the thermoelectric voltage $\Delta V$ generated from the Seebeck effect changes when reversing the direction of the applied temperature gradient [Fig.~\ref{figure1}(a)]. Hence, the difference in voltage $V_\text{USE}=\Delta V_{+}-\Delta V_{-}$ before and after reversing the temperature gradient has been introduced to characterize the USE, where $\Delta V_{+/-}=V_\text{hot}-V_\text{cold}$ represents the Seebeck-induced voltage in an open circuit under the forward ($+$) and backward ($-$) temperature difference $\Delta T$ [Fig.~\ref{figure1}(a)]. $V_\text{hot}$ and $V_\text{cold}$ denote the electrical potential at the hot and cold electrodes, respectively. Additionally, $V_\text{USE}$ has been found to be determined by the quantity $\Delta S$ as $V_\text{USE}=\Delta S \Delta T$ \cite{Yu-2019,Zhu-2023}. Hence, the quantity $\Delta S$, the difference in the Seebeck coefficient between the difference in the Seebeck coefficient between the forward and backward temperature difference, has been introduced to quantify the USE. It has been found that the signal $V_\text{USE}$ comes from the second-order thermal response since the first-order thermal response is eliminated when reversing the temperature gradient \cite{Zhu-2023}. In the experiment, the reversing of the temperature gradient can actually be achieved in the same setup either by switching the laser beam between opposite sides of the sample \cite{Farahman-2015,Shuja-2015} or by applying the direct current to heater electrodes on alternating sides (e.g., prefabricated bilateral electrodes)\cite{Xu-2019}.
USE was firstly predicted in the magnetic topological heterostructures and found to be originated from the asymmetric magnon scattering\cite{Yu-2019}. A subsequent work shows that the USE can also appear in the Rashba 2D electron gas  and is attributed to the Rashba spin-orbit coupling (SOC)\cite{Zhu-2023}. A recent work also exhibits that the Fermi arc discrepancy can also give rise to USE in magnetic Weyl semimetals\cite{Jia-2025}.

The global Rashba SOC is absent, and the spin is degenerate throughout the whole Brillouin zone in $\mathcal{PT}$-symmetric antiferromagnetic materials, which is guaranteed by the global $\mathcal{PT}$ symmetry. However, it has been confirmed that due to the local $\mathcal{P}$ symmetry breaking, the spatially localized Rashba SOC 
in each individual layer (``sectors" ) [Fig.~\ref{figure1}(b)] is nonzero, namely the existence of hidden SOC in the real space\cite{Naka,Ghosh,Wu,Weizhao}. This spatially separated local Rashba SOC in two sectors are opposite, ensuring the global SOC vanishing. Obviously, the hidden spin polarization (HSP) from the local Rashba SOC is odd-distributed in both real and momentum spaces. In addition, the N\'{e}el order is also a space odd-distributed quantity. Recently, it has been reported that the coupling of these two odd-distributed quantities (HSP and N\'{e}el order) yields a superposition effect between the two sectors rather than compensation and gives rise to a macroscopic effect: electrical nonreciprocal transport\cite{Weizhao}.

In this paper, we show that the USE can also exist in the tetragonal $\mathcal{PT}$-symmetric antiferromagnetic materials without the global Rashba SOC but in the presence of the hidden Rashba SOC. The USE is found to stem from the coupling of hidden Rashba SOC and the N\'{e}el vector. Importantly, when the N\'{e}el vector is reversed (180$^{\circ}$ rotation), the sign of the thermal nonlinear coefficient $\chi_{aaa}$ changes. Consequently, the sign of the quantity $\Delta S \propto \chi_{aaa}$ [Eq.~\eqref{S-alpha}] quantifying the USE changes,
offering a new approach to thermally detect the reversal of the  N\'{e}el vector. The relation between the quantity $\Delta S$ and the thermal nonlinear coefficients $\chi_{aaa}$ as a second-order response to $a$-directional temperature gradient is recalled, and the symmetric constraints on the thermal nonlinear coefficients $\chi_{abc}$ are analyzed and determined in Sec.~\ref{There-reve}. We discuss the symmetries of the tetragonal-lattice $\mathcal{PT}$-symmetric AFMs with different orientations of the N\'{e}el vector and also derive its effective tight-binding Hamiltonian based on the generic minimal model (a crinkled quasi-2D model) in Sec.~\ref{SAEM}. The behaviors of USE for the generic minimal model are discussed in Sec.~\ref{RD}. Finally, a conclusion is given in Sec.~\ref{conclusion}.

\section{The concept and symmetric constraints of the unidirectional Seebeck effect}\label{There-reve}
It has been revealed that the USE actually comes from the nonzero thermally driven nonlinear current $j^{(2)}_{a}=\chi_{aaa}(\partial_{a} T)^{2}$ as a second-order response to the temperature gradient $\partial_{a} T$ in the same direction\cite{Zhu-2023}. Since the thermally driven linear current $j^{(1)}_{a}$ will change its direction but $j^{(2)}_{a}$ remains unchanged when reversing the temperature gradient $\partial_{a}T$, the total current $j_{a}=j^{(1)}_{a}+j^{(2)}_{a}$ changes between the cases for the forward $(+)$ and backward $(-)$ temperature difference $\Delta T$, resulting in a change of magnitude of the thermoelectric voltage $\Delta V$ (i.e. $\Delta V_{+}\neq \Delta V_{-} $) in the open circuit [Fig.~\ref{figure1}(a)]. The voltage $\Delta V_{+}$ ($\Delta V_{-}$) represents the Seebeck-induced voltage under the forward (backward) temperature difference. A quantity $\Delta S_{a}$, which describes the difference in the Seebeck coefficients between the cases of the forward and backward temperature gradient, has been phenomenologically introduced to characterize the USE and is given by \cite{Yu-2019}
\begin{equation}
\Delta S_{a}=\frac{2\chi_{aaa}\Delta T}{\sigma_{aa}l}=\frac{2 R_{aa}\chi_{aaa}w\Delta T}{l^{2}}
\label{S-alpha}
\end{equation}
with $\sigma_{aa}$ being the conductivity along $a=(x\,\,\text{or}\,\,y)$ direction, which is aligned with the temperature gradient, $R_{aa}$ denoting the the resistance along $a$ direction, $l (w)$ representing the length (width) of the sample, and $\chi_{aaa}$ (expressed as $\alpha_{aa}^{(2)}$ in the Refs.\cite{Yu-2019,Zhu-2023}) indicating the second-order response coefficient of nonlinear current $j^{(2)}_{a}$ to the temperature gradient $\partial_{a}T$. In the second equality, $R_{aa}=l/(\sigma_{aa}w)$ has been used for the 2D case.
 It should be pointed out the $a$ has been fixed in the $x$-direction in Refs.~\cite{Yu-2019,Zhu-2023}. In this work, we will also consider the case in which $a=y$ since $\chi_{yyy}$ would be crucial for detecting the reversal of the N\'{e}el vector along the $x$-direction (see details in Sec.~\ref{RD}). Accordingly, the difference of the voltage $V_\text{USE}=\Delta V_{+}-\Delta V_{-}$ before and after reversing the  direction of the temperature gradient is found to be \cite{Yu-2019,Zhu-2023}
 \begin{equation}
 V_\text{USE}=\Delta S_{a}\Delta T=\frac{2\chi_{aaa}l(\partial_{a} T)^{2}}{\sigma_{aa}}.
 \label{VUSSD}
 \end{equation}
In obtaining the last equality, $\partial_{a}T\approx {\Delta T}/l$ has been utilized.
Owing to the USE arising from the nonzero $j^{(2)}_{a}$ quantified by the coefficient $\chi_{aaa}$, the signals and behaviors of the USE in the 2D tetragonal $\mathcal{PT}$-symmetric AFMs will be analyzed and discussed based on the calculated coefficient $\chi_{aaa}$.

Based on the semiclassical Boltzmann theory, the nonlinear thermally driven current $j^{(2)}_{a}$ as a second-order temperature gradient can be determined (details can be found in the Appendix \ref{APP-A-NEDF}) and has the following form:
\begin{equation}
\begin{aligned}
j^{(2)}_{a}&=\chi_{abc}\partial_{b}T\partial_{c}T
\label{Ner-coeff}
\end{aligned}
\end{equation}
with
\begin{equation}
\begin{aligned}
\chi_{abc}&=\frac{\tau^{2}e}{ T^{2}}\int[d\mathbf{k}]\frac{(E_\mathbf{k}-E_{f})^{2}}
{m_{ab}} v_{c}\frac{\partial f_{0}}{\partial E_\mathbf{k}},
\end{aligned}
\label{chi-alpa}
\end{equation}
where $\tau$ is relaxation time, $T$ indicates temperature, $\hbar$ means the Planck constant, and  $m_{ab}^{-1}=(1/\hbar)(\partial{v}_{a}/\partial{k_{b}})$ represents the inverse effective mass tensor with group velocity $v_{a}$ in $a$ direction. In time-reversal symmetric ($\mathcal{T}$) or inversion symmetric ($\mathcal{P}$) materials, the nonlinear thermoelectric tensors $\chi_{abc}$ are forbidden. That's because under time reversal or space inversion, the current $j^{(2)}_{a}$ changes sign but $\partial_{b}T\partial_{c}T$ keeps unchanged, hinting that $\chi_{abc}$ also needs to change sign ($\chi_{abc}\xrightarrow{\mathcal{T}\,\,\text{or} \,\mathcal{P}}-\chi_{abc}$), namely $\mathcal{T}$ or $\mathcal{P}$ odd, respectively. The $\mathcal{T}$ ($\mathcal{P}$) odd parity restricts the integrand  of the formula for the coefficient $\chi_{abc}$ in Eq.~\eqref{chi-alpa} to be an odd function of $\mathbf{k}$ and makes  $\chi_{abc}$ forbidden in time-reversal (inversion) symmetric systems.  However, when the materials are $\mathcal{PT}$ symmetric, the nonlinear thermoelectric tensors $\chi_{abc}$ would be nonzero since both current $j^{(2)}_{a}$ and $\partial_{b}T\partial_{c}T$ keep unchanged under $\mathcal{PT}$ symmetry, which makes $\chi_{abc}$ $\mathcal{PT}$ even. In fact, one can also confirm the vanishing of $\chi_{abc}$ in $\mathcal{T}-$ ($\mathcal{P}-$) symmetric systems through exploiting the $\mathbf{k}$-parities given in the Table \ref{tabel-parities} and easily identify the integrand in Eq.~\eqref{chi-alpa} is an odd function of $\mathbf{k}$ when systems is $\mathcal{T}-$ ($\mathcal{P}-$) symmetric.
\begin{table}[tbph]
\centering
\caption{The parities about $\mathbf{k}$ for $\mathcal{P}$ or $\mathcal{T}$ symmetric systems.}
\begin{tabular*}{8 cm}{@{\extracolsep{\fill}}lccc}
 \hline\hline
  \text{Quantities} &$\mathcal{P}$    &$\mathcal{T}$   \\
  \hline
 $E_{\mathbf{k}}$ &\text{even} &\text{even} \\
 $v_{\gamma}$ &\text{odd} &\text{odd}   \\
 $m_{\alpha\beta}$ &\text{even} &\text{even}  \\
 \hline\hline
 \end{tabular*}
 \label{tabel-parities}
 \end{table}

\begin{table*}[tbph]
\centering
\caption{Constraints on in-plane tensor elements of $\chi^{D}$ (Drude terms) from point group symmetries.$``\checkmark"(``\times")$ means the element is symmetry allowed (forbidden). Here $n=2,4,6$ and $m=3,4,6$.}
\begin{centering}
\begin{tabular*}{18 cm}{@{\extracolsep{\fill}}lcccccccccccccccccccccc}
\hline \hline
 \thead{}& \thead{$\mathcal{PT},\sigma_{z}$\\$C_{3}^{z},C_{2}^{z}\mathcal{T}$}& \thead{$\mathcal{P},\mathcal{T}$\\$C_{n}^{z},S_{4,6}^{z},S_{6}^{x,y}$}& \thead{$C_{m}^{z}\mathcal{T},C_{3}^{x}\mathcal{T}$\\$C_{3,6}^{y}\mathcal{T},\sigma_{z}\mathcal{T},S_{4,6}^{z}\mathcal{T}$}& \thead{$C_{n}^{x},\sigma_{y}$\\$C_{2}^{y}\mathcal{T},\sigma_{x}\mathcal{T},S_{4,6}^{x}\mathcal{T}$}& \thead{$C_{n}^{y},\sigma_{x}$\\$C_{2}^{x}\mathcal{T},\sigma_{y}\mathcal{T},S_{4,6}^{y}\mathcal{T}$}& \thead{$S_{4}^{x},C_{4}^{x}\mathcal{T}$}& \thead{$S_{4}^{y},C_{4}^{y}\mathcal{T}$}& \thead{$C_{3}^{x}$}& \thead{$C_{3}^{y}$}& \thead{$C_{6}^{x}\mathcal{T}$}& \\
 \hline
 $\chi_{yxx}$& $\checkmark$& $\times$& $\times$& $\times$& $\checkmark$& $\times$& $\checkmark$& $\times$& $\checkmark$& $\times$& \\
 $\chi_{xyy}$& $\checkmark$& $\times$& $\times$& $\checkmark$& $\times$& $\checkmark$& $\times$& $\checkmark$& $\times$& $\times$&\\
 $\chi_{xxx}$& $\checkmark$& $\times$& $\times$& $\checkmark$& $\times$& $\times$& $\checkmark$& $\checkmark$& $\checkmark$& $\times$&\\
 $\chi_{yyy}$& $\checkmark$& $\times$& $\times$& $\times$& $\checkmark$& $\times$& $\times$& $\checkmark$& $\checkmark$& $\checkmark$&\\
 \hline \hline
\end{tabular*}
\par\end{centering}
\label{table-symmetry}
\end{table*}

In addition to the $\mathcal{P}$ and $\mathcal{T}$ symmetries, other crystalline symmetries also impose constraints on $\chi_{abc}$, a third-rank pseudotensor, as
\begin{equation}
\chi_{abc}=\eta_{T} O_{aa^{\prime}}O_{bb^{\prime}}
O_{cc^{\prime}}\chi_{a^{\prime}b^{\prime}c^{\prime}},
\end{equation}
where $O\equiv \mathcal{R}\,(\mathcal{R}\mathcal{T})$ is a nomagnetic (magnetic) point group operation, and $\eta_{T}=\pm 1$ is connected with the $\mathcal{T}$-odd character of $\chi_{abc}$: $\eta_{T}=+1$ for the nonprimed operation (i.e. operation $\mathcal{R}$) and $\eta_{T}=-1$ for the primed operations, namely the magnetic symmetry operations of the form $\mathcal{R}\mathcal{T}$. The obtained constraints are summarized in Table \ref{table-symmetry}.

\section{Symmetries analysis and effective model for the tetragonal $\mathcal{PT}$ antiferromagnets}\label{SAEM}
Since the nonlinear thermoelectric response coefficient $\chi_{aaa}$ is allowed by the $\mathcal{PT}$ symmetry, one kind of the candidate lattices to observe the USE is the tetragonal $\mathcal{PT}$-symmetric AFMs composed of two sublattices $A$ and $B$ [Figs.~\ref{figure1}(b) and (c)]. The atoms $A$ and $B$ in the unit cell with opposite in-plane magnetization are symmetric in respect to the $\mathcal{PT}$ center. Therefore, two sublattices belong to two sectors, $\alpha$ and $\beta$, respectively. The nomagnetic space group of the tetragonal lattice is $P4/nmm$.
When the N\'{e}el vector is along the $x$-direction ($y$-direction), the space group $P4/nmm$ is reduced to the magnetic space group (MSG) $Pmm^{\prime}n$ ($Pm^{\prime}mn$) for the tetragonal lattice $\mathcal{PT}$-symmetric AFMs, and the corresponding magnetic point group (MPG) is the MPG $mm^{\prime}m$ ($m^{\prime}mm$). When the N\'{e}el vector lies in the $xy$-plane but not in the $x$-direction and $y$-direction, the MSG and  MPG will be further reduced to $P2^{\prime}_{1}/m$ and  $2^{\prime}/m$, respectively. The group elements of MSGs ($Pmm^{\prime}n$, $Pm^{\prime}mn$, $P2^{\prime}_{1}/m$) and MPGs ($mm^{\prime}m$, $m^{\prime}mm$, $2^{\prime}/m$ ) are listed in Table {\ref{Table3}}. According to the Tables \ref{table-symmetry} and {\ref{Table3}}, the symmetric operations  ($C^{y}_{2}$, $\sigma_{x}$, $C^{x}_{2}\mathcal{T}$, $\sigma_{y}T$) of MPG $mm^{\prime}m$ will enforce $\chi_{xyy}=\chi_{xxx}=0$ for $\hat{\mathbf{n}}//[100]$. On the contrary, for $\hat{\mathbf{n}}//[010]$, one can have $\chi_{yxx}=\chi_{yyy}=0$, which are restricted by the symmetries ($C^{x}_{2}$, $\sigma_{y}$, $C^{y}_{2}\mathcal{T}$, $\sigma_{x}T$) of MPG $m^{\prime}mm$. Additionally, for the MPG $2^{\prime}/m$, only the symmetries $C^{z}_{2}T$, $m_{z}$ and $\mathcal{PT}$ are survived, and, consequently, all the nonlinear thermal coefficients [$\chi_{xyy}$, $\chi_{xxx}$, $\chi_{yxx}$, $\chi_{xxx}$] are allowed. The nonzero $\chi_{xxx}$ ($\chi_{yyy}$) means that the USE can exist  and be observed in tetragonal lattice $\mathcal{PT}$-symmetric AFMs when the temperature gradient is applied along the $x$ ($y$) direction.
\begin{table*}[tbph]
\centering
\caption{The group elements for the magnetic space groups ($Pmm^{\prime}n$, $Pm^{\prime}mn$, $P2^{\prime}_{1}/m$) and magnetic point groups ($mm^{\prime}m$, $m^{\prime}mm$, $2^{\prime}/m$ )}
\begin{centering}
\begin{tabular*}{18 cm}{@{\extracolsep{\fill}}c|cc}
\hline\hline
 \thead{} &\thead{International notation} &\thead{Element} \\
 \hline
 \thead{\multirow{5}*{Magnetic Space Group}} &\thead{$Pmm^{\prime}n$}
  &\thead{$I,\mathcal{S}_{y}=\{C^{y}_{2}|0,1/2,0\},\mathcal{G}_{x}=\{\sigma_{x}|1/2,0,0\},
  \mathcal{G}_{z}=\{\sigma_{z}|1/2,1/2,0\}$\\$\mathcal{PT},\mathcal{S}_{x}\mathcal{T}=\{C^{x}_{2}|1/2,0,0\}\mathcal{T},
  \mathcal{S}_{z}T=\{C^{z}_{2}|1/2,1/2,0\}T,\mathcal{G}_{y}T=\{\sigma_{y}|0,1/2,0\}T$} \\
\cmidrule{2-3}
                                     &\thead{$Pm^{\prime}mn$} &\thead{$I,\mathcal{S}_{x}=\{C^{x}_{2}|1/2,0,0\},\mathcal{G}_{y}
                                     =\{\sigma_{y}|0,1/2,0\},
                                     \mathcal{G}_{z}=\{\sigma_{z}|1/2,1/2,0\}$\\$\mathcal{PT},
                                     \mathcal{S}_{y}\mathcal{T}=\{C^{y}_{2}|0,1/2,0\}\mathcal{T},
                                     \mathcal{S}_{z}\mathcal{T}=\{C^{z}_{2}|1/2,1/2,0\}\mathcal{T},\mathcal{G}_{x}
                                     \mathcal{T}                                   =\{\sigma_{x}|1/2,0,0\}\mathcal{T}$}\\
\cmidrule{2-3}
                                     &\thead{$P2^{\prime}_{1}/m$}  &\thead{$I,\mathcal{PT},\mathcal{G}_{z}=\{\sigma_{z}|1/2,1/2,0\},
                                     \mathcal{S}_{z}\mathcal{T}=\{C^{z}_{2}|1/2,1/2,0\}\mathcal{T}$}\\
\hline
\thead{\multirow{5}*{Magnetic Point Group}} &\thead{$mm^{\prime}m$} &\thead{$I,C_{2y},\sigma_{x},\sigma_{z}$\\$\mathcal{PT},C^{x}_{2}\mathcal{T},C^{z}_{2}\mathcal{T},\sigma_{y}\mathcal{T}$}   \\
\cmidrule{2-3}
                                    &\thead{$m^{\prime}mm$}&\thead{$I,C^{x}_{2},\sigma_{y},\sigma_{z}$\\
                                    $\mathcal{PT},C^{y}_{2}\mathcal{T},C^{z}_{2}\mathcal{T},\sigma_{x}\mathcal{T}$}\\
\cmidrule{2-3}
                                    &\thead{$2^{\prime}/m$}  &\thead{$I,\mathcal{PT},\sigma_{z},C^{z}_{2}\mathcal{T}$}\\
\hline\hline
\end{tabular*}
\par\end{centering}
\label{Table3}
\end{table*}

We consider a generic minimal model for the 2D tetragonal lattice $\mathcal{PT}$-symmetric AFMs (including the 2D antiferromagnetic tetragonal
CuMnAs\cite{Hong-2025,Sepehrina-2024,smejkal-2017}), in which the two sublattices $A$ and $B$ (with one orbital per atom) form a stack of the crinkled quasi-2D square lattices [Figs.~\ref{figure1}(b) and (c)]. The tight-binding Hamiltonian for the minimal model is
\begin{equation}
\begin{aligned}
H_{0}&=-t\sum_{i\in A}\sum_{\bm{\delta}_{1},\sigma}\hat{a}^{+}_{i,\sigma}b_{i+\bm{\delta}_{1},\sigma}-
t^{\prime}\sum_{\bm{\delta}_{2},\sigma}
\left(\sum_{i\in A}\hat{a}^{+}_{i,\sigma}a_{i+\bm{\delta}_{2},\sigma}\right.\\
&\left.+\sum_{i\in B}\hat{b}^{+}_{i,\sigma}
b_{i+\bm{\delta}_{2},\sigma}\right)+h.c.,
\end{aligned}
\end{equation}
where the first (second) term represents the intersublattice $A$-$B$ (intrasublattice $A$-$A$ and $B$-$B$) hopping term with the first (second) nearest-neighbor hopping strength $t$ ($t^{\prime}$), $i$ indicates the lattice sites, $\hat{a}^{+}_{i,\sigma}$ ($\hat{a}_{i,\sigma}$) and $\hat{b}^{+}_{i,\sigma}$ ($\hat{b}_{i,\sigma}$) are the electron creation (annihilation) operators for spin $\sigma=\uparrow,\downarrow$  on sublattice $A$ and sublattice $B$ at the $i$-th site, respectively, and $\bm{\delta_{1}}$ ($\bm{\delta_{2}}$) means a 2D displacement vector to the first- (second-)  nearest neighbor atoms. For the investigated crinkled quasi-2D square lattice antiferromagnet [Fig.~\ref{figure1}(c)], one has $\bm{\delta_{1}}=(0.5,\pm0.5)a$ and $(-0.5,\pm0.5)a$ to the nearest neighbors and $\bm{\delta_{2}}=(\pm a,0)$ and $(0,\pm a)$ to all the next nearest neighbors, where $a$ denotes the lattice constant. When only considering the nearest neighbor exchange interaction, the AFM exchange interaction $H_\text{AFM}$ can  be expressed as
 \begin{equation}
 H_\text{AFM}=J_{n}\sum_{i}\left(\hat{a}^{+}_{i}\hat{a}_{i}-\hat{b}^{+}_{i}\hat{b}_{i}\right)
 \bm{\sigma}\cdot\mathbf{n},
 \end{equation}
where $J_{n}$ characterizes the overall strength of the AFM exchange interaction and $\mathbf{n}$ indicates the N\'{e}el vector, which is defined as the difference of the magnetic moments between atom A and B in the unit cell. For the hidden SOC, we consider the Rashba SOC \cite{Cheng-2012}, which is given by
\begin{equation}
\begin{aligned}
H_\text{R}&=i\lambda_\text{R}\sum_{i\in A}\sum_{\bm{\delta}_{2}\sigma,\sigma^{\prime}}
a^{+}_{i,\sigma}\left(\bm{\hat{\sigma}}\times\hat{\bm{\delta}}_{2}\right)^{z}_{\sigma\sigma^{\prime}}
a^{+}_{i+\bm{\delta}_{2},\sigma^{\prime}}\\
&-i\lambda_\text{R}\sum_{i\in B}\sum_{\bm{\delta}_{2}\sigma,\sigma^{\prime}}
b^{+}_{i,\sigma}\left(\bm{\hat{\sigma}}\times\hat{\bm{\delta}}_{2}\right)^{z}_{\sigma\sigma^{\prime}}
b^{+}_{i+\bm{\delta}_{2},\sigma^{\prime}}.
\end{aligned}
\end{equation}
One might notice that $\lambda_\text{R}$ is for  the $A$ sublattices but $-\lambda_\text{R}$ is for
the $B$ sublattices. That's because the direction of the effective dipole or electric field in the $\alpha$-sector (atom $A$ located in) is opposite for that in the $\beta$-sector (atom $B$ located in) owing to the $\mathcal{PT}$ symmetry. By performing Fourier transformations, the effective Hamiltonian $H_\mathbf{k}$ in basis ($\{|A\rangle,|B\rangle\}\otimes\{|\uparrow\rangle,|\downarrow\rangle\}$) is found to be
\begin{equation}
\begin{aligned}
H_\mathbf{k}=&-2t\tau_{x}\cos\frac{k_{x}}{2}\cos\frac{k_{y}}{2}-t^{\prime}\left(\cos k_{x}
+\cos k_{y}\right)\\
&+\tau_{z}\bm{\sigma}\cdot\left(\mathbf{\lambda}_\mathbf{k}+J_{n}\mathbf{n}\right),
\label{Hamil}
\end{aligned}
\end{equation}
where $\bm{\tau}$ and $\bm{\sigma}$ are Pauli matrices representing
sublattice and spin degrees of freedom, respectively, $\bm{\lambda}=\lambda(-\sin k_{y},\sin k_{x})$ is the effective magnetic field induced by SOC with $\lambda$ representing the strength of second-neighbor SOC, $J_{n}$ represents the AFM exchange coupling strength, $\mathbf{n}$ is the N\'{e}el vector, and the wave vector $\mathbf{k}$ is in units of the inverse lattice constant. Owing to the presence of global $\mathcal{PT}$ symmetry, the spin is degenerate throughout the whole Brillouin zone (BZ), thus, there are the two sets of spin-degenerate bands with the eigenvalues
\begin{equation}
\begin{aligned}
E_{\mathbf{k}}(\pm)=&-t^{\prime}(\cos k_{x}+\cos k_{y})\pm\left[4t^{2}\cos^{2}\frac{k_{x}}{2}\cos^{2}\frac{k_{x}}{2}\right.\\
&\left.\!+\!\left(J_{n}n_{x}-\lambda\sin k_{y}\right)^{2}\!+\!\left(J_{n}n_{y}\!+\!\lambda\sin k_{y}\right)^{2}\!+\!J_{n}^{2}n_{z}^{2}\right]^{\frac{1}{2}},
\end{aligned}
\label{Emm}
\end{equation}
where the ``$+$" (``$-$") in  $E_{\mathbf{k}}(\pm)$ represent the conduction (valence) band, respectively. Although the energy band is spin-degenerate in the momentum space, the spin polarization manifests locally on the atom sites in each sector when breaking the local $\mathcal{PT}$ symmetry. The HSP, which characterizes local spin polarization in real space, is defined as projecting the Bloch wave function $|\psi_{n\mathbf{k}}\rangle$ of the double (spin) degenerate band onto a specific sector (for example, sector $\alpha$) and has the form as $\mathbf{P}^{\alpha}_\mathbf{k}=\sum_{i\in\alpha}\sum_{n\in\mathbb{N}}\langle\psi_{n\mathbf{k}}|
 \mathbf{\sigma}\bigotimes| i\rangle\langle i|\psi_{n\mathbf{k}}\rangle$, \cite{Weizhao,Ying-2020} with $\mathbf{k}$, $\mathbf{\sigma}$, and $|i\rangle$ representing the wavevector, Pauli operator, and the localized orbitals belonging to sector $\alpha$, respectively. $\mathbb{N}$ means the summation over two spin-degenerate bands. Therefore, the corresponding HSPs for Eq.~(\ref{Hamil}) are $\mathbf{P}_\mathbf{k}(\pm)=(\mathbf{\lambda}_{\mathbf{k}}+J_{n}\mathbf{n})/
 \sqrt{|\mathbf{\lambda}_{\mathbf{k}}+J_{n}\mathbf{n}|^{2}+|t_\mathbf{k}|^{2}}$ with $t_\mathbf{k}=2t\cos\frac{k_{x}}{2}\cos\frac{k_{y}}{2}$ for simplicity. Obviously, the HSPs $\mathbf{P}_\mathbf{k}(\pm)$ for the 2D tetragonal $\mathcal{PT}$-symmetric AFMs are dependent on the intersublattice A-B hopping $t$ but independent of the intrasublattice hopping $t^{\prime}$.
 It should be pointed out that the effective Hamiltonian [Eq.~\eqref{Hamil}] is derived based on the 2D case but can also serve as a minimal model for the special three-dimensional tetragonal $\mathcal{PT}$-symmetric AFMs (e.g., the thin-film CuMnAs\cite{smejkal-2017}), in which the distance between the  quasi-2D planes is larger than the first and second nearest-neighbor distances within  each quasi-2D planes, making it reasonable to neglect the coupling between the quasi-2D planes to capture the essential physics of the system.

\begin{figure}
\centering
\includegraphics[width=1.0\linewidth,clip]{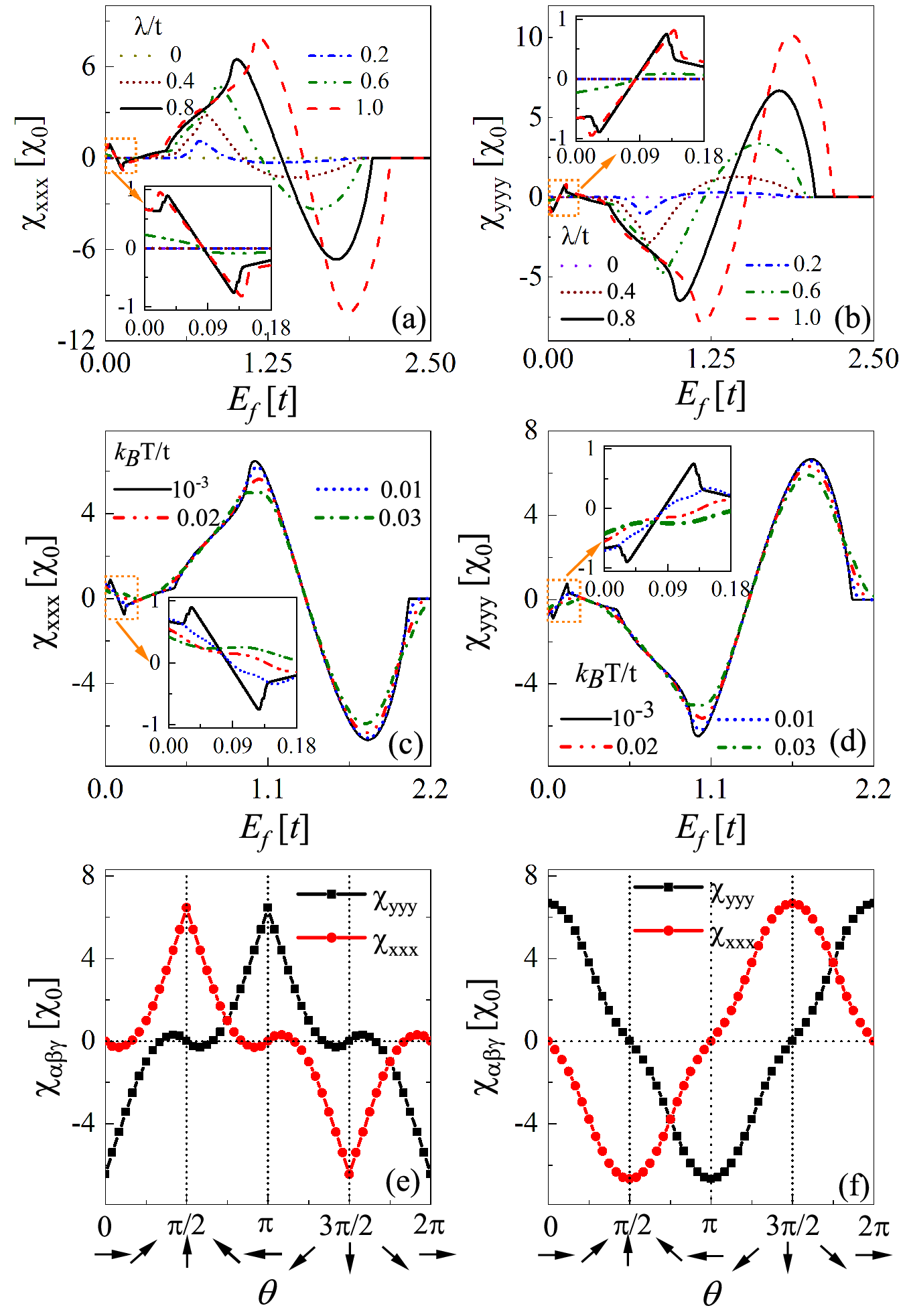}
\caption{The coefficients $\chi_{xxx}$ and $\chi_{yyy}$ as a function of Fermi energy $E_{f}$ for different Rashba strengths $\lambda$ [(a) and (b)] and for different temperatures $T$ [(c) and (d)]. The dependence $\chi_{aaa}$ ($a=x\,\text{or}\,\ y$) on the orientation of the  N\'{e}el vector [i.e., $\theta$] [(e) and (f)]. The arrows on the bottoms indicate the direction of the N\'{e}el vector, and $\theta=0$ corresponds to the N\'{e}el is along the x-axis direction. $k_{B}T/t$ is taken to be $1\times 10^{-3}$ in (a), (b), (e) and (f). $\lambda=0.8t$ is fixed at (c)-(f). $E_{f}=1.0 t$ is taken in (e) and $E_{f}=1.78 t$  is fixed in (f). The typical scale $\chi_{0}$ is defined as $\chi_{0}=2\tau^{2}ek_{B}^{2}t/(4\pi^{2}\hbar^{3}k_{0})$. Other parameters are used as follows: $j_{n}=0.6t$ and $t^{\prime}=0.08t$.\cite{Hong-2025,smejkal-2017}}
\label{figure2}
\end{figure}

\section{Results and discussion} \label{RD}
Recent studies\cite{Yu-2019,Zhu-2023} have shown that both asymmetric magnon scattering and asymmetric energy dispersion (band asymmetry) can give rise to the USE. In the absence of magnetic impurity, however, the contribution of asymmetric magnon scattering to USE vanishes. Therefore, the prerequisite for the generation of nonzero USE in pure tetragonal lattice $\mathcal{PT}$-symmetric AFMs becomes the band asymmetry with respect to the opposite vector, namely $E_{\mathbf{k}}\neq E_{-\mathbf{k}}$.  According to Eq.~(\ref{Emm}), the magnitude of $\Delta E_\mathbf{k}=E_{\mathbf{k}}-E_{-\mathbf{k}}$, which quantifies the band asymmetry, is found to be
\begin{equation}
\begin{aligned}
\Delta E_{\mathbf{k}}(\pm)&=\frac{\pm4\mathbf{\lambda}_{\mathbf{k}}\cdot J_{n}\mathbf{n} }{\sqrt{|\mathbf{\lambda}_{\mathbf{k}}+J_{n}\mathbf{n}|^{2}+|t_\mathbf{k}|^{2}}+
\sqrt{|\mathbf{\lambda}_{\mathbf{k}}-J_{n}\mathbf{n}|^{2}+|t_\mathbf{k}|^{2}}}\\
&=\frac{\pm4P_{\mathbf{k}}^\text{ave}\mathbf{\lambda}_{\mathbf{k}}\cdot J_{n}\mathbf{n}}{|\mathbf{\lambda}_{\mathbf{k}}+ J_{n}\mathbf{n}|+|\mathbf{\lambda}_{\mathbf{k}}- J_{n}\mathbf{n}|},
\end{aligned}
\label{detal2}
\end{equation}
where $P_{\mathbf{k}}^\text{ave}=\{w|\mathbf{P}_{\mathbf{k}}^{\alpha}|+(1-w)|\mathbf{P}_{\mathbf{k}}^{\alpha}|\}$ is the standardized average of the HSP with weight $w=\sqrt{|\mathbf{\lambda}_{\mathbf{k}}+ J_{n}\mathbf{n}|^{2}+|t_{\mathbf{k}}|^{2}}/\{\sqrt{|\mathbf{\lambda}_{\mathbf{k}}+ J_{n}\mathbf{n}|^{2}+|t_{\mathbf{k}}|^{2}}+\sqrt{|\mathbf{\lambda}_{\mathbf{k}}- J_{n}\mathbf{n}|^{2}+|t_{\mathbf{k}}|^{2}}\}$. According to Eq.~(\ref{detal2}), one can easily observe that the $\Delta E_{\mathbf{k}}$ vanishes when $\mathbf{\lambda}_{\mathbf{k}}\perp\mathbf{n}$, and $\Delta E_{\mathbf{k}}$
is proportional to the HSP for other situations
 in which the $\mathbf{\lambda}_\mathbf{k}$ is not perpendicular to $\mathbf{n}$. Besides, the $\Delta E_{\mathbf{k}}$ also disappears, and the energy dispersion becomes symmetric when either the strength $\lambda$ of Rashba SOC or the AFM exchange coupling strength $J_{n}$ is zero. Hence, one can conclude that
  in the 2D tetragonal lattice $\mathcal{PT}$-symmetric AFMs, the joint effect of the magnetic order (N\'{e}el vector), SOC, and the local symmetry breaking as the key reason leads to the band asymmetry, which further results in a nonlinear USE.

To further investigate and analyze the behaviours of $\chi_{xxx}$ ($\chi_{yyy}$), we use the following parameters: $J_{n}=0.6t$ and $t^{\prime}=0.08t$, which are consistent to the typical strengths of the 2D antiferromagnetic tetragonal CuMnAs. The Rashba strength $\lambda$ can be modulated through applying strain \cite{Liu} or applying an electric field (gate voltage) perpendicular to the sample surface to regulate the energy band structure \cite{Caviglia,Gong}. Here, we take $\lambda$ in the range of  $[0,1.0t]$\cite{Smejkal}. When  $\lambda>J_{n}$, there are two Dirac points (DPs) $D_{1}$ and $D_{2}$ [Fig.~\ref{figure1}(e)] located at $[\pi,\arcsin(J_{n}/\lambda)]~ ([-\arcsin(J_{n}/\lambda),\pi])$  and $[\pi,\pi-\arcsin(J_{n}/\lambda))]~([\arcsin(J_{n}/\lambda)-\pi,\pi])$ for the N\'{e}el vector $\mathbf{n}\parallel[100]~ ([010])$, respectively. The corresponding energies for DPs $D_{1}$ and $D_{2}$ are $t^{\prime}-t^{\prime}\sqrt{1-J^{2}_{n}/\lambda^{2}}$ and $t^{\prime}+t^{\prime}\sqrt{1-J^{2}_{n}/\lambda^{2}} $, respectively.  However, when $\lambda<J_{n}$ or the N\'{e}el vector is in plane but not along [100] ([010]), the energy gap opens and the DPs disappear. A typical scale $\chi_{0}=2\tau^{2}ek_{B}^{2}t/(4\pi^{2}\hbar^{3}k_{0})$ (where $k_{0}=10^{10}~ \mathrm{m}^{-1}$ and the factor 2 is accounted for the spin degeneracy) is also introduced to characterize the nonlinear thermal response coefficients $\chi_{xxx}$ $(\chi_{yyy})$. $\chi_{0}$ is equal to $21.14~\mathrm{nA\cdot n m/K^{2}}$ when $t=1.0~\mathrm{eV}$ (a value for the 2D antiferromagnetic tetragonal CuMnAs\cite{Hong-2025,Sepehrina-2024}) and  $\tau=10^{-12}~\mathrm{s}$ \cite{Gomoany-2016} (a typical value for AFMs). The relaxation time $\tau$ is not well known for the tetragonal CuMnAs, here the typical value for AFMs is utilized for the estimation.

Figures \ref{figure2}(a) and (b) show the coefficients $\chi_{xxx}$ and $\chi_{yyy}$ as a function of Fermi energy $E_{f}$ for different Rashba-SOC strengths $\lambda$, respectively.
 One can easily observe that $\chi_{xxx}$ [Fig.~\ref{figure2}(a)] has the same magnitude but opposite sign  to the coefficient $\chi_{yyy}$ [Fig.~\ref{figure2}(b)]. Hence, we will focus on the discussion of the dependence of $\chi_{xxx}$ on the $E_{f}$ and $\lambda$ for simplicity.
As expected, the coefficient $\chi_{xxx}$ becomes zero when $\lambda=0$, and, consequently, the USE vanishes. For a tiny upshift of $E_{f}$ in the electron doping, $\chi_{xxx}$ displays a small positive peak followed by a negative peak when $\lambda>J_{n}=0.6t$ [the insets in Figs.~\ref{figure2} (a) and (c)]. It's found that these two peaks appear close to the two DPs. 
However, when $\lambda\leq J_{n}=0.6t$ or at high temperature $k_{B}T$, these peaks near the band edge disappear owing to the disappearance of DPs for $\lambda\leq J_{n}=0.6t$ and the temperature broadening effect in high temperature. In addition to the peaks near the band edge, one would also observe a larger positive peak accompanied by a negative peak in the high doping as long as $\lambda\neq0$. This larger positive peak is found to appear along the $XM$ (or $M^{\prime}X^{\prime}$) (for example, $E_{f}=1.0t$ for $\lambda=0.8t$). 
One can shift peak positions to higher Fermi levels and enlarge the peak value through increasing the Rashba SOC strength. Besides, the magnitudes of the two larger peaks for $\chi_{xxx}$ slightly decrease with increasing of temperature, and the positions of the positive (negative) peaks shift toward the higher (lower) Fermi levels [Figs.~\ref{figure2}(c) and (d)].

Figure \ref{figure2}(c) [(d)] illustrates the dependence of the nonlinear thermal coefficient $\chi_{xxx}$ [$\chi_{yyy}$] on the orientation of the N\'{e}el vector $\mathbf{n}$. One can observe that $\chi_{xxx}=0$ ($\chi_{yyy}=0$) when the N\'{e}el vector $\mathbf{n}$ is along the $\pm x$ ($\pm y$ ) direction, i.e., $\theta=0$\, or $\pi$  ($\theta=\pi/2$ or $3\pi/2$), which is consistent to the symmetric constraints discussed in Sec.~\ref{SAEM}. Additionally, when $\mathbf{n}$ is along $a=(x,y)$ direction, the $|\chi_{a_{\perp}a_{\perp}a_{\perp}}|$ reaches its maximum, where $a_{\perp}$ indicates the coordinate axis orthogonal to the axis $a$ in the 2D plane.
Importantly, both $\chi_{xxx}$ and $\chi_{yyy}$ exhibit a $2\pi$ periodicity when $\mathbf{n}$ rotates along the $x$-$y$ plane and change sign when reversing the direction of the N\'{e}el vector $\mathbf{n}$, namely satisfying the $\mathcal{T}$-odd constraint: $\chi_{aaa}(\theta)=-\chi_{aaa}(\theta+\pi)$ (where $a=x,y$). According to Eq.~(\ref{VUSSD}) one can also identify that the sign of the difference of voltage $V_\text{USE}$, which stems from and characterizes the USE, depends on the sign of $\chi_{aaa}$.
Therefore, one can recognize $180^{\circ}$ reversal of the N\'{e}el vector $\mathbf{n}$ through measuring the sign-changing of $V_\text{USE}$, offering a new thermal approach to detect $180^{\circ}$ reversal of the N\'{e}el vector. Particularly, when the temperature gradient is applied along the $x$-direction ($y$-direction), the sign of the USE, which is determined by $\chi_{xxx}$ ($\chi_{yyy}$) [Eq.~\eqref{Ner-coeff}], can be utilized to recognize the reversal of $\mathbf{n}$ in the $\pm x$ direction  ($\pm y$ direction).

To numerically estimate the order of magnitude of  $V_\text{USE}~[=2\chi_{xxx}l(\partial_{x}T)^{2}/\sigma_{xx}]$ generated from the USE, we take $\chi_{xxx}=6.45\chi_{0}$ for $E_{f}=1.0t$ and $\lambda=0.8t$ [Fig.~\ref{figure2} (a)]. The conductivity $\sigma_{xx}=7.54\times10^{-4}~\Omega^{-1}=3.11 e^{2}/\hbar$ for the 2D antiferromagnetic tetragonal  CuMnAs is estimated from the bulk resistivity $\rho$ through $\sigma_{xx}=h/\rho_{xx}$ with $h$ indicating the thickness of the 2D sample. The bulk resistivity $\rho$ in the bulk antiferromagnetic tetragonal CuMnAs can range from 90 to 160~$\mathrm{\mu\Omega~cm}$. We take $\rho_{xx}=90~\mathrm{\mu\Omega~cm}$ and $h=c=6.318~{\AA}$  ($c$ denotes the unit-cell thickness) \cite{Wadlely-2013} for estimation, respectively. In the experiment, the temperature gradient $\partial_{x}T$ can already reach 1.5$~K\mu m^{-1}$.\cite{Xu-2019} Therefore, the differential voltage $V_{\text{USE}}$ can reach 0.1$~m V$ at $l=120~\mu m$, which is well within the experimentally measurable voltage range since the voltage detection sensitivities as low as 0.1 $\mu V$ have been achieved\cite {Uchida-2008}.

\section{conclusion} \label{conclusion}
We have analyzed the symmetric constraints on the USE. It's found that the USE disappears in $\mathcal{P}$- or $\mathcal{T}$-symmetric materials but can exist in $\mathcal{PT}$-symmetric materials. Specially, we investigate USE in tetragonal $\mathcal{PT}$ antiferromagnetic materials and analyze the dependence of USE on the orientation of the N\'{e}el vector based on the symmetric constraints. Our results show that when $\hat{\mathbf{n}}//[100]$ ([010]), the temperature gradient should be applied along $\pm x$ ($\pm y$) to achieve the nonzero USE. To further investigate the behaviors of USE in 2D tetragonal $\mathcal{PT}$-symmetric antiferromagnetic materials, the tight-binding effective Hamiltonian has been deduced for the generic minimal model (a crinkled quasi-2D model ). We find that the nonzero USE attributes to the joint effect of the hidden Rashba SOC and the N\'{e}el vector in 2D tetragonal $\mathcal{PT}$-symmetric antiferromagnetic materials without the global Rashba SOC. Importantly, when reversing the N\'{e}el vector, the sign of USE changes, which would offer a new thermal approach to detect the reversal of N\'{e}el vector. It should be pointed out that the discussed phenomenon here would be observed in various $\mathcal{PT}$-symmetric antiferromagnetic materials, including monolayer V$_2$Se$_2$O,\cite{HaYang-2021} monolayer FeSe\cite{Wu-2024,Qiao-2020}, and tetragonal CuMnAs\cite{Hong-2025,smejkal-2017,Sepehrina-2024}, and can be applied to recognize their reversal of the N\'{e}el vector in plane.
\section{acknowledgements}
This work is supported by the National Natural Science Foundation of China (Grant No. 12004107
), the National Science Foundation of Hunan, China (Grant No. 2023JJ30118), and the Fundamental Research Funds for the Central Universities.
\\

\appendix
\section{The derivation of the nonlinear thermal response coefficients $\chi_{abc}$}
\label{APP-A-NEDF}
In the absence of the external electric field and magnetic field, the Boltzmann equation for the electron distribution within the relaxation time approximation is
\begin{equation}
f-f_{0}=-\tau \frac{\partial f}{\partial r_{a}}\cdot v_{a},
\label{App-A-1}
\end{equation}
with $f_{0}=1/(\textrm{exp}[\frac{\epsilon_{\mathbf{k}}-E_{f}}{k_{B}T}]+1)$ indicating the equilibrium Fermi distribution, $\tau$ denoting the relaxation time, and $r_{a}$ and $v_{a}$ representing the $a$ components of coordinate position and velocity of electrons, respectively. Since we are interested in the response up to second order in temperature gradient, we have the nonequilibrium distribution function $f\approx f_{0}+\delta f_{1}+\delta f_{2}$ with the terms $f_{n}$ understood to vanish as $(\partial T/\partial r_{a})^{n}$. Assuming the uniform  temperature gradient in the system, namely $\partial_{ab}T =0$, the nonequilibrium distribution function $\delta f_{1}\left(\mathbf{k}\right)$  and  $\delta f_{2}\left(\mathbf{k}\right)$
 as the first-order and second-order response to the temperature gradient have been determined as \cite{Zhu-2023}, respectively,
\begin{equation}
\begin{aligned}
\delta f_{1}&\!=\!\frac{\tau}{T\hbar}(\epsilon_\mathbf{k}-E_{f})\frac{\partial f_{0}}{\partial k_{a}}\partial_{a}T,\\
\delta f_{2}&\!=\!\tau^{2}\left[2\hbar v_{b}\frac{\partial f_{0}}{\partial k_{a}}+\left(\epsilon_\mathbf{k}\!-\!E_{f}\right)\frac{\partial ^{2}f_{0}}{\partial k_{a}\partial k_{b}}\right]\!\frac{\epsilon_\mathbf{k}-E_{f}}{\hbar^{2}T^{2}}\partial_{a}T\partial_{b}T.
\end{aligned}
\label{App-A-F12F}
\end{equation}
It should be pointed out that if the temperature gradient is applied in a single direction, one can have $a=b$ in Eq. (\ref{App-A-F12F}). Since the charge current in the $a$-direction is $j_{a}=-e\int[d\mathbf{k}]v_{a} f\left(\mathbf{r},\mathbf{k}\right)$, one can find the $a$-component of nonlinear current $j^{(2)}_{a}$ in the second-order temperature gradient has the following form.
\begin{equation}
\begin{aligned}
j^{(2)}_{a}&=-e\int[d\textbf{k}]v_{a}\delta f_{2}\\
&=-\frac{\tau^{2}e}{\hbar^{2}T^{2}}\int[d\mathbf{k}]v_{a}(\epsilon_\mathbf{k}-E_{f})\left[2\hbar v_{b}\frac{\partial f_{0}}{\partial k_{a}}\right.\\
&\left.+\left(\epsilon_\mathbf{k}-E_{f}\right)\frac{\partial ^{2}f_{0}}{\partial k_{a}\partial k_{b}}\right]\partial_{a}T\partial_{b}T.
\label{App-A-j2}
\end{aligned}
\end{equation}
In obtaining the second line in Eq.~\eqref{App-A-j2},  the determined formula for $\delta f_{2}$ in Eq.~(\ref{App-A-F12F}) has been substituted into the first line. Combining the definition of $j^{(2)}_{a}=\chi_{abc}\partial_{b}T\partial_{c}T$ with Eq.~\eqref{App-A-j2}, one would easily verify that
\begin{widetext}
\begin{equation}
\begin{aligned}
\chi_{abc}&=-\frac{\tau^{2}e}{\hbar^{2}T^{2}}\int[d\mathbf{k}]v_{a}
(\epsilon_\mathbf{k}-E_{f})\left[2\hbar v_{b}\frac{\partial f_{0}}{\partial k_{c}}+\left(\epsilon_\mathbf{k}-E_{f}\right)\frac{\partial ^{2}f_{0}}{\partial k_{b}\partial k_{c}}\right].\\
&=-\frac{\tau^{2}e}{\hbar^{2}T^{2}}\int[d\mathbf{k}]v_{\alpha}
(\epsilon_\mathbf{k}-E_{f})2\hbar^{2} v_{b}v_{c}\frac{\partial f_{0}}{\partial \epsilon_\mathbf{k}}-\frac{\tau^{2}e}{\hbar^{2}T^{2}}\int[d\mathbf{k}]v_{a}
(\epsilon_\mathbf{k}-E_{f})^{2}\frac{\partial }{\partial k_{b}}(\hbar v_{c}\frac{\partial f_{0}}{\partial \epsilon_\mathbf{k}})\\
&=-\frac{\tau^{2}e}{\hbar^{2}T^{2}}\int[d\mathbf{k}]v_{a}
(\epsilon_\mathbf{k}-E_{f})2\hbar^{2} v_{b}v_{c}\frac{\partial f_{0}}{\partial \epsilon_\mathbf{k}}+\frac{\tau^{2}e}{\hbar^{2}T^{2}}\int[d\mathbf{k}]\frac{\partial }{\partial k_{b}}\left[
 v_{a}(\epsilon_\mathbf{k}-E_{f})^{2}\right]\hbar v_{c}\frac{\partial f_{0}}{\partial \epsilon_\mathbf{k}}\\
 &=-\frac{\tau^{2}e}{\hbar^{2}T^{2}}\int[d\mathbf{k}]v_{a}
(\epsilon_\mathbf{k}-E_{f})2\hbar^{2} v_{b}v_{c}\frac{\partial f_{0}}{\partial \epsilon_\mathbf{k}}+\frac{\tau^{2}e}{\hbar T^{2}}\int[d\mathbf{k}]\left[\frac{(\epsilon_\mathbf{k}-E_{f})^{2}}
{m_{ab}}+2v_{a}v_{b}\left(\epsilon_{\mathbf{k}}-E_{f}\right)
\right]\hbar v_{c}\frac{\partial f_{0}}{\partial \epsilon_\mathbf{k}}\\
&=\frac{\tau^{2}e}{ T^{2}}\int[d\mathbf{k}]\frac{(\epsilon_\mathbf{k}-E_{f})^{2}}
{ m_{ab}} v_{c}\frac{\partial f_{0}}{\partial \epsilon_\mathbf{k}},
\label{App-A-coff}
\end{aligned}
\end{equation}
\end{widetext}
where $m_{ab}^{-1}=(1/\hbar)(\partial{v}_{a}/\partial{k_{b}})$.  The relation $v_{a}={\partial \epsilon_\mathbf{k}}/({\hbar\partial k_{a}})$ has been used to obtain the second line{\color{blue}{,}} and the partial integration has been applied to obtain the third line.

\end{document}